\documentstyle[12pt]{article}
\begin{document}
\begin{titlepage}
\title{The Role of the $\Delta(1920)$ Resonance
for Kaon Production in Heavy Ion Collisions}
\author{K. Tsushima\thanks{Supported by DFG under contract No.
Fa 67/14-1},
S.W. Huang\thanks{Supported by GSI under contract No. 06 T\"U 736},
and Amand Faessler\\ \\
Institut f\"ur Theoretische Physik, Universit\"at T\"ubingen\\
Auf der Morgenstelle 14, D-72076 T\"ubingen, F. R. Germany}
\date{}
\maketitle
\begin{abstract}
The long mean free path of $K^+$ mesons in nuclear matter makes
this particle a suitable messenger for the dynamics of
nucleus-nucleus reactions at intermediate energies (100 MeV to 3
GeV per nucleon). A prerequisite for this is the knowledge of
the elementary production cross sections $\pi N \rightarrow \Sigma K$.
Here these cross sections are studied for the first time with
the explicite inclusion of the relevant baryon resonances up to
2 GeV as intermediate states. The baryon resonances
--  $N(1710)\, I(J^P) = \frac{1}{2} (\frac{1}{2}^+),\,
N(1720)\, \frac{1}{2} (\frac{3}{2}^+)$
and $\Delta(1920)\, \frac{3}{2} (\frac{3}{2}^+)\,$ --
are taken into account coherently in the calculations of the
$\pi N \rightarrow \Sigma K$ process.
(We refer to this model as the `resonance model'.) Also
$K^*(892)\frac{1}{2}(1^-)$ vector meson exchange is included.
It is shown that the total cross sections for
different channels of the $\pi N \rightarrow \Sigma K$
reactions, i.e.
$\pi^+ p \rightarrow \Sigma^+ K^+$,
$\pi^- p \rightarrow \Sigma^- K^+$,
$\pi^+ n \rightarrow \Sigma^0 K^+$
($\pi^- p \rightarrow \Sigma^- K^+$) and
$\pi^0 p \rightarrow \Sigma^0 K^+$ differ not only by absolute
values but also by their energy dependence.
This shape differences are due to the mixture of the isospin
$I = 3/2$ $\Delta(1920)$ with isospin $I = 1/2$ nucleon resonances.
However, this $I = 3/2$ resonance does not give a contribution to
the $\pi N \rightarrow \Lambda K$ reactions.
So the shapes of the total cross sections
$\pi N \rightarrow \Lambda K$
for different isospin projections are the same.
In spite of this, such cross sections averaged over
different isospin projections in the same multiplet are often
used for studies of kaon production in heavy-ion collisions.
Here we give an explicit separate parametrization of the cross
sections for each channel $\pi N \rightarrow \Sigma K$,
which are calculated in this work.
The total cross section for
$\pi^- p \rightarrow \Sigma^- K^+$
which runs through different $I = 3/2$ and $I = 1/2$
resonances can be well reproduced by the present model.
\end{abstract}
\end{titlepage}

%
Because kaons ($K^+$) have a long mean free path inside the
nucleus, one believes that they are good messengers to provide
information about the high density and temperature phase of the heavy ion
collisions \cite{sch,abb}. Furthermore, they can be also sensitive
probes for the nuclear equation of state (EOS) \cite{aic}.
For these reasons, many studies of kaon
production in heavy ion collisions have been performed both
theoretically and experimentally \cite{sch}-\cite{cug}.

In most theoretical studies of the kaon production in heavy ion collisions
by either the Vlasov-Uehling-Uhlenbeck approach (VUU)
\cite{vuu}, or by ``quantum'' molecular dynamics (QMD)
\cite{qmd,li}, however, the kaon elementary production cross sections
in free space parametrized by J. Randrup and C. M. Ko \cite{ran}
for $B_1 B_2 \rightarrow B_3 Y K$, and by
J. Cugnon and R. M. Lombard \cite{cug} for $\pi N \rightarrow Y
K$ are widely used. Here $B$ stands for either the nucleon or
the $\Delta(1232)$, $Y$ stands for either the $\Sigma$ or the
$\Lambda$, respectively. Although there exists a relatively
long history for kaon production studies,
not many investigations have been performed for the
elementary process \cite{fer}.

On the other hand, due to recent high precision data,
we have now rich and accurate
informations on the baryon resonances, on quantum numbers,
on decay modes, on decay rates and so on \cite{par,paro}. Thus, it
is now adequate and necessary to incorporate these resonances in
calculations of the elementary cross section for kaon production.
There are some works which include such baryon
resonances in kaon production studies \cite{bro,ko2}.
In these, simple parametrizations by Breit-Wigner
forms are given. Since the $\pi N \rightarrow \Lambda K$
amplitudes arising from the relevant resonances are not calculated,
the interferences among these resonances is not included. Furthermore,
the $\pi N \rightarrow \Sigma K$ reactions are not investigated
in these publications.

The first purpose of the present article is to present
model calculations of
\newline $\pi N \rightarrow \Sigma K$
at intermediate energies, which take into account the nucleon-
$(I = 1/2)$ and the delta $(I = 3/2)$ resonances up to 2 GeV in
the intermediate states. By intermediate energies we mean that
the center-of-mass energy $\sqrt{s}$ of the $\pi N$
ranges from the threshold of kaon production to about 3 GeV.
Kaon production by pions on nucleons are so-called secondary
processes in heavy ion collisions, which are known to give
about a 30 \% contribution
to kaon production in the intermediate or low energy region \cite{hal}.

The energy dependence of the total cross sections and the peak positions
( e.g. $\pi^- p \rightarrow \Lambda K^0$ reaction \cite{bal})
show that the s-channel resonance processes
given in fig. 1 are dominant for the
$\pi N \rightarrow Y K$ reactions.
( See also the experimental data given in fig. 3 (a). )

According to the compilation of the ``Review of Particle Properties''
\cite{par,paro}, the resonances
$N(1710)\, I(J^P) = \frac{1}{2}(\frac{1}{2}^+),\,\,
 N(1720)\, \frac{1}{2}(\frac{3}{2}^+)$ and
$\Delta(1920)\, \frac{3}{2}(\frac{3}{2}^+)$
give the main s-channel contributions to the reactions
$\pi N \rightarrow \Sigma K$. Although it is expected to be
small (and it will be shown so) we consider the lightest $K^*(892)
\frac{1}{2} (1^-)$ vector meson for t-channel $K^*$ exchange contributions.

The second purpose of this article is to focus on the role
of the $\Delta(1920)$ resonance for the $\pi N \rightarrow \Sigma K$
reaction, since this resonance has isospin $I = 3/2$ and contributes
differently from those of the isospin $I = 1/2$ nucleon resonances.
Because the experimental data \cite{par,paro} show that
this resonance does not decay to $\Lambda K$,
we will not discuss the $\pi N \rightarrow \Lambda K$ reactions
in the present article. A detailed study including the
$\Lambda K$ channel will be reported elsewhere \cite{hua}.

In order to see the different isospin structure of the processes
represented by (a), (b), (c) and (d) in fig. 1, we consider
the following schematic structures in isospin space,
\begin{eqnarray}
(a):\qquad
&( \bar{K} \overrightarrow{\bar \Sigma} \cdot \vec\tau N(1710) )
\,\, ( \bar{N}(1710) \vec\tau \cdot \vec\phi N ), \label{isoa} \\
(b):\qquad
&( \bar{K} \overrightarrow{\bar \Sigma} \cdot \vec\tau N(1720) )
\,\, ( \bar{N}(1720) \vec\tau \cdot \vec\phi N ), \label{isob} \\
(c):\qquad
&( \bar{K} \overrightarrow{\bar \Sigma}
\cdot {\overrightarrow{\cal I}}^\dagger
\Delta(1920) ) \,\,
( \bar{\Delta}(1920) \overrightarrow{\cal I} \cdot \vec\phi N ),
\label{isoc} \\
(d):\qquad
&( \bar{K} \vec\tau K^*(892) \cdot \vec\phi\, ) \,\,
( \bar{K}^*(892) \overrightarrow{\bar \Sigma} \cdot \vec\tau N ),
\label{isod}
\end{eqnarray}
where $\vec{\cal I}$ is the transition operator defined by
\begin{eqnarray*}
\overrightarrow{\cal I}_{Mm} &= \displaystyle{\sum_{\ell=\pm1,0}}
(1 \ell \frac{1}{2} m | \frac{3}{2} M)
\hat{e}^*_{\ell},
\end{eqnarray*}
and $\vec \tau$ are the Pauli matrices.
$N, N(1710), N(1720)$ and $\Delta(1920)$
stand for the fields of $N(938)$, $N(1710), N(1720)$ and
$\Delta(1920)$ resonances. They are expressed by
$\bar{N} = \left( \bar{p}, \bar{n} \right)$,
similarly to the nucleon resonances, and
$\bar{\Delta}(1920) = ( \bar{\Delta}(1920)^{++},
\bar{\Delta}(1920)^+, \bar{\Delta}(1920)^0,
\bar{\Delta}(1920)^- )$.
The fields appearing in eqs. (\ref{isoa}) - (\ref{isoc}) are
related to the physical representations as follows:
$K^T = \left( K^+, K^0 \right),\,\,
\bar{K} = \left( K^-, \bar{K^0} \right),\,\,
K^*(892)^T = \left( K^*(892)^+, K^*(892)^0 \right),\,\,
\bar{K}^*(892) = \left( K^*(892)^-, \bar{K}^*(892)^0 \right),\,\,
\pi^{\pm} = \frac{1}{\sqrt{2}} (\phi_1 \mp i \phi_2),\,\,
\pi^0 = \phi_3,\,\,
\Sigma^{\pm} = \frac{1}{\sqrt{2}} (\Sigma_1 \mp i \Sigma_2),\,\,
\Sigma^0 = \Sigma_3\,\,$,
where the superscript $T$ means
transposition operation.
The pseudoscalar meson fields
are defined as annihilating (creating) the physical particle
(anti-particle) states. $SU(2)$ isospin symmetry is
assumed for each doublet or multiplet.

Without factors arising from isospin structure, we define the amplitudes
${\cal M}_a, {\cal M}_b, {\cal M}_c$ and
${\cal M}_d$ corresponding to each diagram
(a), (b), (c) and (d) given in fig. 1.
Then each amplitude contributing to the
$\pi N \rightarrow \Sigma K$ reactions
(which will be refered as `different channels') is given by,
\begin{eqnarray}
{\cal M}_c+2{\cal M}_d\,\, {\rm for}
&\pi^+ p \rightarrow \Delta(1920)^{++}, K^*(892)^0 \rightarrow
\Sigma^+ K^+ \nonumber \\
{\rm and} &\pi^- n \rightarrow \Delta(1920)^-, K^*(892)^+
\rightarrow \Sigma^- K^0, \label{ampc} \\
\nonumber \\
2({\cal M}_a+{\cal M}_b+\frac{1}{6}{\cal M}_c)\,\, {\rm for}
&\pi^- p \rightarrow
n(1710), n(1720), \Delta^0(1920) \rightarrow
\Sigma^- K^+ \nonumber \\
{\rm and} &\pi^+ n \rightarrow
p(1710), p(1720), \Delta^+(1920) \rightarrow
\Sigma^+ K^0, \label{iso1} \\
\nonumber \\
\sqrt{2} ({\cal M}_a+{\cal M}_b-\frac{1}{3}{\cal M}_c-{\cal
M}_d) & \nonumber \\
{\rm for} &\pi^+ n \rightarrow
p(1710), p(1720), \Delta(1920)^+, K^*(892)^0 \rightarrow
\Sigma^0 K^+ \,\, \nonumber\\
{\rm and} &\pi^0 p \rightarrow
p(1710), p(1720), \Delta(1920)^+, K^*(892)^0 \rightarrow
\Sigma^+ K^0, \,\, \label{iso2} \\
\nonumber \\
-\sqrt{2} ({\cal M}_a+{\cal M}_b-\frac{1}{3}{\cal M}_c-{\cal
M}_d) & \nonumber \\
{\rm for} &\pi^0 n \rightarrow
n(1710), n(1720), \Delta(1920)^0, K^*(892)^+ \rightarrow
\Sigma^- K^+ \,\, \nonumber \\
{\rm and} &\pi^- p \rightarrow
n(1710), n(1720), \Delta(1920)^0, K^*(892)^+ \rightarrow
\Sigma^0 K^0, \,\, \label{iso3} \\
\nonumber \\
{\cal M}_a+{\cal M}_b+\frac{2}{3}{\cal M}_c+{\cal M}_d &
\nonumber \\
{\rm for} &\pi^0 p \rightarrow
p(1710), p(1720), \Delta(1920)^+, K^*(892)^+ \rightarrow
\Sigma^0 K^+ \,\, \nonumber\\
{\rm and} &\pi^0 n \rightarrow
n(1710), n(1720), \Delta(1920)^+, K^*(892)^0 \rightarrow
\Sigma^0 K^0. \,\, \label{iso4}
\end{eqnarray}
Here the intermediate resonance states are explicitely
written between the two arrows pointing to the right.
Note that $I = 1/2$ t-channel $K^*$ meson exchanges
do not contribute to the channel given by eq. (\ref{iso1}) due
to the isospin structure.
{}From eqs. (\ref{ampc}) - (\ref{iso4}) one can see that
the $\Delta(1920)$(represented by the amplitude ${\cal M}_c$)
 contributes to different channels in
different ways than the other $N(1710)$ and $N(1720)$ resonances
(represented by the amplitudes ${\cal M}_a$ and ${\cal M}_b$)
and $K^*(892)$ exchange (represented by ${\cal M}_d$).

This means that the $\Delta(1920)$ resonance modifies the
energy dependence of the cross section for the different
channels of $\pi N \rightarrow \Sigma K$ reactions.
(Except for the reactions given in eq. (\ref{iso2})
and eq. (\ref{iso3}).)
This is completely different from the
$\pi N \rightarrow \Lambda K$ reactions, where only the isospin
$I = 1/2$ nucleon resonances can contribute in s-channel and
there is no t-channel $K^*$ exchange contributions.

Before proceeding to the specific model calculations,
we can roughly estimate the $\Delta(1920)$ resonance
contributions to the total cross sections of each different
channel defined by eqs.(\ref{iso1}) - (\ref{iso4}) neglecting
$K^*(892)$ exchange contribution.
The experimental total cross
section $\sigma(\pi^+ p) (\propto |{\cal M}_c^2|)$
for the $\pi^+ p \rightarrow \Sigma^+ K^+$ channel has a
peak with a value of about 0.7 mb located around $\sqrt{s} = 1.92$
GeV as seen from fig. 3 (a). We assume the
$\Delta(1920)$ contribution is dominant, due to the experimental
branching ratios \cite{par,paro}, and
the fact that only charge $2 e$ intermediate states are allowed.
Then the incoherent contributions of this $\Delta(1920)$ resonance to each
total cross section at this peak position may be estimated as follows:
$\frac{1}{9}|{\cal M}_c|^2\propto \frac{1}{9}\sigma(\pi^+ p) \simeq 0.08$ mb
for the channels defined by eq. (\ref{iso1}),
$\frac{2}{9}|{\cal M}_c|^2\propto \frac{2}{9}\sigma(\pi^+ p) \simeq 0.16$ mb
for the channels defined by
eqs. (\ref{iso2}) and (\ref{iso3}), and
$\frac{4}{9}|{\cal M}_c|^2 \propto \frac{4}{9}\sigma(\pi^+ p) \simeq 0.32$ mb
for the channels defined by eq. (\ref{iso4}), respectively.
Thus, this $\Delta(1920)$ resonance can give sizable contributions
to each channel.
{}From this estimate we expect that the cross sections for
the different channels in the $\pi N \rightarrow \Sigma K$
reactions will be different from each other not only in absolute
values but also in its energy dependence.

In order to illustrate this argument more clearly,
we perform here model calculations for these cross sections.
Effective interaction Lagrangians for the processes represented by fig. 1
are used:
\begin{equation}
{\cal L}_{\pi N N(1710)} =
-ig_{\pi N N(1710)}
\left( \bar{N}(1710) \gamma_5 \vec\tau N \cdot \vec\phi
+ \bar{N} \vec\tau \gamma_5 N(1710) \cdot \vec\phi\,\, \right),
\label{lagfirst}
\end{equation}
\begin{equation}
{\cal L}_{\pi N N(1720)} =
\frac{g_{\pi N N(1720)}}{m_\pi}
\left( \bar{N}^\mu(1720) \vec\tau N \cdot \partial_\mu \vec\phi
+ \bar{N} \vec\tau N^\mu(1720) \cdot \partial_\mu \vec\phi \,
\right),
\end{equation}
\begin{equation}
{\cal L}_{\pi N \Delta(1920)} =
\frac{g_{\pi N \Delta(1920)}}{m_\pi}
\left( \bar{\Delta}^\mu(1920) \overrightarrow{\cal I} N \cdot
\partial_\mu \vec\phi + \bar{N} {\overrightarrow{\cal I}}^\dagger
\Delta^\mu(1920) \cdot \partial_\mu \vec\phi \, \right),
\end{equation}
\begin{equation}
{\cal L}_{K \Sigma N(1710)} =
-ig_{K \Sigma N(1710)}
\left( \bar{N}(1710) \gamma_5 \vec\tau \cdot \overrightarrow\Sigma K
+ \bar{K} \overrightarrow{\bar \Sigma} \cdot \vec\tau
\gamma_5 N(1710) \right),
\end{equation}
\begin{equation}
{\cal L}_{K \Sigma N(1720)} =
\frac{g_{K \Sigma N(1720)}}{m_K}
\left( \bar{N}^\mu(1720) \vec\tau \cdot \overrightarrow\Sigma
\partial_\mu K + (\partial_\mu \bar{K}) \overrightarrow{\bar \Sigma}
\cdot \vec\tau N^\mu(1720) \right),
\end{equation}
\begin{equation}
{\cal L}_{K \Sigma \Delta(1920)} =
\frac{g_{K \Sigma \Delta(1920)}}{m_K}
\left( \bar{\Delta}^\mu(1920) \overrightarrow{\cal I}
\cdot \overrightarrow\Sigma \partial_\mu K
+ (\partial_\mu \bar{K}) \overrightarrow{\bar \Sigma} \cdot
{\overrightarrow{\cal I}}^\dagger \Delta^\mu(1920) \right),
\end{equation}
\begin{equation}
{\cal L}_{K^*(892) \Sigma N} = - g_{K^*(892) \Sigma N}
\left( \bar{N} \gamma^\mu \vec\tau \cdot \overrightarrow\Sigma K^*_\mu(892)
+ \frac{\xi}{m_N + m_\Sigma} \bar{N} \sigma^{\mu \nu}
\vec\tau \cdot \overrightarrow\Sigma \partial_\mu K^*_\nu(892)
+ {\rm h. c.} \right), \label{ksn}
\end{equation}
\vspace{1mm}
\begin{equation}
{\cal L}_{K^*(892) K \pi} = i f_{K^*(892) K \pi}
\left( \bar{K} \vec\tau K^*_\mu(892) \cdot
\partial^\mu \vec\phi
- (\partial^\mu \bar{K}) \vec\tau K^*_\mu(892) \cdot \vec\phi \,\right)
+ {\rm h. c.}  ,
\label{laglast}
\end{equation}
\\
where $\xi$ is the ratio of the tensor coupling constant to the
vector coupling constant.
The spin 3/2 Rarita-Schwinger fields
$\psi^\mu = N^\mu(1720),\, \Delta^\mu(1920)$ with mass $m$
satisfy the set of quations \cite{tak},
\begin{eqnarray}
 & ( i \gamma \cdot \partial - m ) \psi^\mu = 0, \\
 & \gamma_\mu \psi^\mu = 0, \\
 & \partial_\mu \psi^\mu = 0.
\end{eqnarray}
There are several problems in
order to treat spin 3/2 massive particles and
spin 1/2 and 3/2 resonances consistently.
They are discussed in great detail in refs. \cite{ben}, and \cite{wil,pec,aur}.
The problems discussed in these references are, for example, connected
with the uniqueness of the free Lagrangian for the massive spin 3/2 particles,
with the propagator having a definite spin 3/2 projection,
with the unitarity due to the widths inserted into the
propagators of the resonances, and so on.  There seems to be
according to Benmerrouche et al. \cite{ben} also a problem with
the constraint to the off shell propagator of spin 3/2 particles suggested
by Peccei \cite{pec}.
Although keeping these discussions in mind we must use a
phenomenological specific form for the propagators of the
resonances.
We use for the propagators $S_F(p)$ of the spin 1/2 and
$G^{\mu \nu}(p)$ of the spin 3/2 resonances,
\begin{equation}
S_F(p) = \frac{\gamma \cdot p + m}{p^2 - m^2 + im\Gamma^{full}}\,,
\end{equation}
\begin{equation}
G^{\mu \nu}(p) = \frac{P^{\mu \nu}(p)}{p^2 - m^2 +
im\Gamma^{full}}\,,
\end{equation}
with
\begin{equation}
P^{\mu \nu}(p) = - (\gamma \cdot p + m)
\left[ g^{\mu \nu} - \frac{1}{3} \gamma^\mu \gamma^\nu
- \frac{1}{3 m}( \gamma^\mu p^\nu - \gamma^\nu p^\mu)
- \frac{2}{3 m^2} p^\mu p^\nu \right], \label{pmunu}
\end{equation}
\newline
where $m$ and $\Gamma^{full}$ stand for the mass and the full decay
width of the corresponding resonance.

Without factors arising from isospin space, the explicit amplitudes
${\cal M}_a$, ${\cal M}_b$, ${\cal M}_c$ and ${\cal M}_d$
corresponding diagrams (a), (b), (c) and (d) depicted in fig. 1 are
given as follows:
\begin{equation}
{\cal M}_a = \frac{- g_{\pi N N(1710)}g_{K \Sigma N(1710)}}
{p^2 - m_{N(1710)}^2 + i m_{N(1710)}
\Gamma_{N(1710)}^{full}} \enspace
\bar{u}_\Sigma (p_\Sigma)\, \gamma_5\, (\gamma \cdot p + m_{N(1710)})\,
\gamma_5\, u_N(p_N), \label{ma}
\end{equation}
\begin{equation}
{\cal M}_b = \frac{g_{\pi N N(1720)}g_{K \Sigma N(1720)}}{m_\pi m_K}\,
\frac{{p_K}_\mu {p_\pi}_\nu}
{p^2 - m_{N(1720)}^2 + i m_{N(1720)}
\Gamma_{N(1710)}^{full}} \enspace
\bar{u}_\Sigma (p_\Sigma)\, P_{N(1720)}^{\mu \nu}(p)\,
u_N(p_N), \label{mb}
\end{equation}
\begin{equation}
{\cal M}_c = \frac{g_{\pi N \Delta(1920)}g_{K \Sigma \Delta(1920)}}
{m_\pi m_K}\,
\frac{{p_K}_\mu {p_\pi}_\nu}
{p^2 - m_{\Delta(1920)}^2 + i m_{\Delta(1920)}
\Gamma_{\Delta(1920)}^{full}} \enspace
\bar{u}_\Sigma (p_\Sigma)\, P_{\Delta(1920)}^{\mu \nu}(p)\, u_N(p_N),
\label{mc}
\end{equation}
$$
{\cal M}_d =
\frac{f_{K^*(892) K \pi} g_{K^*(892) \Sigma N}}
{(p_\Sigma - p_N)^2 - m_{K^*(892)}^2} \enspace
\bar{u}_\Sigma (p_\Sigma)\,
\left[ \gamma_\mu - i \frac{\xi}{ m_N + m_\Sigma }
\sigma_{\alpha \mu} (p_\Sigma - p_N)^\alpha \right] \hspace{3cm}
$$
\begin{equation}
\hspace{5cm} \cdot (p_\pi + p_K)_\nu \left( g^{\mu \nu} -
\frac{(p_\Sigma - p_N)^\mu (p_\Sigma - p_N)^\nu}{m_{K^*(892)}^2} \right)\,
u_N(p_N),
\label{md}
\end{equation}
where $u_N(p_N)$ and $u_\Sigma(p_\Sigma)$ are the spinors of the
nucleon and the $\Sigma$, with the four momenta
$p_N$ and $p_\Sigma$, respectively. (See fig. 1.)

We also need to determine the coupling
constants appearing in the Lagrangians given by
eqs. (\ref{lagfirst}) - (\ref{laglast}).
In order to include the finite size effects of hadrons,
form factors (denoted by $F(q)$ and $F_{K^*(892)K\pi}(q)$ below) will
be introduced.
These form factors must be multiplied to each vertex of the interaction.
Thus, the absolute values of the coupling constants except for
$g_{K^*(892) \Sigma N}$ are determined from the branching
ratios in the rest frame of the resonances:
$$
\Gamma(N(1710) \rightarrow B P) =
3 \frac{g^2_{P B N(1710)} F^2(q(m_{N(1710)},m_B,m_P))}{4\pi}\hspace{9em}
$$
\begin{equation}
\hspace{18em} \cdot \frac{(E_B - m_B)}{m_{N(1710)}}
q(m_{N(1710)},m_B,m_P), \label{ratio}
\end{equation}
\begin{equation}
\Gamma(B^* \rightarrow B P) =
D \frac{g^2_{P B B^*} F^2(q(m_{B^*},m_B,m_P))}{12\pi}
\frac{(E_B + m_B)}{m_{B^*} m_P^2}
q^3(m_{B^*},m_B,m_P), \label{gamma}
\end{equation}
$$
\Gamma(K^*(892) \rightarrow K \pi) =
3 \frac{f^2_{K^*(892) K \pi}F^2_{K^*(892) K \pi}
(q(m_{K^*(892)},m_K,m_\pi))}{4\pi}\hspace{6em}
$$
\begin{equation}
\hspace{18em} \cdot
\frac{2}{3 m^2_{K^*(892)}} q^3(m_{K^*(892)},m_K,m_\pi), \label{kgamma}
\end{equation}
with
\begin{equation}
F(q) = \frac{\Lambda_C^2}{\Lambda_C^2 + q^2},\qquad
F_{K^*(892) K \pi}(q)
= C q \exp\left( - \beta q^2 \right),
\label{form}
\end{equation}
\begin{equation}
q(x,m_B,m_P) = \frac{1}{2 x}
\left[ (x^2 - (m_B + m_P)^2)\,(x^2 - (m_B - m_P)^2) \right]^{1/2},
\label{qdeff}
\end{equation}
\newline
where $E_B = \sqrt{m_B^2 + \vec{p}\hspace{1mm}^2_B}\,\, (q =
|\vec{p}_B|)$,
$B^*$ stands for either the $N(1720)$ or the $\Delta(1920)$ resonance,
$B$ stands for either the $N(938)$ or the $\Sigma$, and $P$ stands for
either the $\pi$ or the $K$, respectively.
Also the corresponding masses are represented by
$m_{B^*}$, $m_B$ and $m_P$.
In the above, $F(q)$ is the form factor with a cut off parameter $\Lambda_C$,
and  $D = 3$ and $D = 1$ should be assigned to the $N(1720)$ and
the $\Delta(1920)$, respectively. The $K^*(892) K \pi$ vertex
form factor is taken from ref. \cite{gob}. In evaluating the cross
sections in the center-of-mass frame of the $\pi N$ system, each coupling
constant
$g_{PBB^*}$, $g_{K^*(892)\Sigma N}$ and $f_{K^*(892)K\pi}$
appearing in the eqs. (\ref{ma}) - (\ref{md}) must be replaced by
$g_{PBB^*} \rightarrow g_{PBB^*}F(q(\sqrt{s},m_B,m_P))$,
$g_{K^*(892)\Sigma N} \rightarrow g_{K^*(892)\Sigma N}
F( (\vec{q}_f-\vec{q}_i) )$
and
$f_{K^*(892)K\pi} \rightarrow f_{K^*(892)K\pi}
F_{K^*(892)K\pi}( \frac{1}{2}(\vec{q}_f-\vec{q}_i) )$,
with $\vec{p}_B = - \vec{p}_P$, $q = |\vec{p}_B|$,
$|\vec{q}_f| = q(\sqrt{s},m_\Sigma,m_K)$,
$|\vec{q}_i| = q(\sqrt{s},m_N,m_\pi)$ and $s$ being
the Mandelstam variable. (See also the caption of fig. 2.)

The cut off parameters used for the numerical calculations are:
$\Lambda_C = 0.8$ GeV for the $N(1710)$ and the $N(1720)$ resonances,
$\Lambda_C = 0.5$ GeV for the $\Delta(1920)$ resonance and
$\Lambda_C = 1.2$ GeV for the $K^*(892) \Sigma N$ vertex, respectively.
The parameters $C$ and $\beta$ in $F_{K^*(892) K \pi}$ are
$C = 2.72$ fm and $\beta = 8.88 \times 10^{-3}$ fm$^2$ used in
ref. \cite{gob}.

We first discuss the $\pi^+ p \rightarrow \Sigma^+ K^+$ channel
displayed in fig 3 (a). The experimental data are given with error
bars \cite{bal}. Note that only the $\Delta(1920)$ contributes
to this reaction in the s-channel. (See eq. (\ref{ampc}).)
The dashed line stands for the result using the coupling
constants determined from the branching ratios
and eqs. (\ref{gamma}) - (\ref{qdeff}), and the fitted value for
$g_{K^*(892) \Sigma N}$ which will be explained later.
This shows a large underestimation -- of about a factor 12 --
if branching ratio data are used to determine the coupling
constants for the baryon resonances. The $K^*(892)$ exchange
contribution to this total cross section
gives an almost constant background as a function of energy, and thus
must be small so that it is
consistend with the experimental data.
The total cross section is mainly due to
the $\Delta(1920)$ resonance in the
present model, thus the two coupling constants relevant for this
resonance (see eq. (\ref{gamma}) and Table 1) are rescaled in order to fit
the data in fig. 3 (a). The coupling constant $g_{K^*(892) \Sigma N}$ is
chosen to fit the tail of this cross section with zero tensor
coupling for $K^*(892) \Sigma N$ interaction, i.e.
$\xi = 0$ in eq. (\ref{ksn}),
and the rescaled coupling constants for $\Delta(1920)$
are used. This value for $g_{K^*(892) \Sigma N}$
is used for the results represented by the dashed line in fig. 3 (a).
The result using the rescaled coupling constants and the value
$g_{K^*(892) \Sigma N}$ fitted as described above is represented by the
solid line in fig. 3 (a).
This set of coupling constants will be also used later.

In the following calculations, two different sets of the coupling
constants will be used.
For set 1 we use the rescaled coupling
constants for the $\Delta(1920)$ resonance and the fitted
coupling constant $g_{K^*(892) \Sigma N}$. All the other
coupling constants are determined from the branching ratios.
The results obtained using the coupling
constants determined fully by the experimental branching ratios
and the same value of $g_{K^*(892) \Sigma N}$ as case 1 will be refered
as set 2.
Explicite values of the coupling constants obtained and the experimental
branching ratios are listed in table 1.

Hereafter, we will discuss the $K^+$ production channels only.
Corresponding arguments for the $K^0$ production channels
are valid as seen from eqs. (\ref{ampc}) - (\ref{iso4}).

The calculated total cross sections for the other
channels are displayed in figs. 3(b) - 3(d).
The contributions of the interference
terms appearing in these channels are included as follows:
Since the branching ratios determine only the square of the
coupling constants, all eight possible relative signs among
the interference terms are evaluated.
Then that sign combination which reproduces
the experimental data best for $\pi^- p \rightarrow \Sigma^- K^+$
is selected. (See fig. 3 (b).) After fixing the relative signs among the
interference terms for this reaction, all other channels are
evaluated with this fixed relative signs.

First we discuss the global features of figs. 3(b) - 3(d).
It is clear that the total cross sections of different channels
show a different energy dependence, for both set 1 and set 2.
This different energy dependence is due to the $\Delta(1920)$
resonance contributions as already mentioned in the text.
Without the $\Delta(1920)$ resonance contributions, they all
have the same energy dependence but different absolute values.
Thus, this isospin $I = 3/2$ $\Delta(1920)$ resonance
differentiates the energy dependences of the total
cross sections for the different channels in the
$\pi N \rightarrow \Sigma K$ reactions.

Next, it is interesting to look at the $\pi^- p \rightarrow \Sigma^- K^+$
channel in fig. 3 (b). The experimental data are also given
with error bars \cite{bal}.
The parameters of set 1 can reproduce the second peak of
the experimental data around the total center-of-mass energy
$\sqrt{s} = 1.95$ GeV.
But the parameters of set 2 fail to reproduce this second peak.
This second peak described correctly by the parameters of set 1
is again the consequence of the isospin $I = 3/2$ $\Delta(1920)$ resonance
contribution. Thus, this fact shows the superiority of the
parameters of the set 1 for the present study.
The success of set 1 and the failure of set 2 to explain
the $\pi^+ p \rightarrow \Sigma^+ K^+$ and
$\pi^- p \rightarrow \Sigma^- K^+$
data simultaneously might indicate that there
exist other resonances with masses around 1.9 GeV and isospin
$I = 3/2$ which can give large enough contributions to reproduce
the $\pi^+ p \rightarrow \Sigma^+ K^+$ total cross section,
and also give the second peak of the reaction
$\pi^- p \rightarrow \Sigma^- K^+$.

In figs. 3(c) and 3(d), the calculated total cross sections for the
$\pi^+ n \rightarrow \Sigma^0 K^+$ and
$\pi^0 n \rightarrow \Sigma^- K^+$, and
$\pi^0 p \rightarrow \Sigma^0 K^+$ are shown.
Set 1 results for $\pi^0 p \rightarrow \Sigma^0 K^+$ should be reliable
prediction.

Since the energy-dependent form of the decay widths is not unique,
energy independent decay widths are used in the
propagators for the present study.
To see the effects of energy-dependent decay widths
in the propagators for the resonances,
some calculations have also been done using an
energy dependent decay width
$\Gamma(s) = \Gamma^{full}
\left( \frac{q(\sqrt{s},m_B,m_P)}{q(m,m_B,m_P)} \right)^{2L+1}$,
where $L$ is the orbital angular momentum quantum number of the
corresponding resonance.
(See also eq. (\ref{qdeff}) and fig. 2.) It turned out that this
energy dependence of the decay width does not give significant effects
on the energy dependence of the cross sections.

For kaon production in heavy ion collisions, we give parametrizations
of the total cross sections $\sigma(\pi N \rightarrow \Sigma K)$
for each separate channel based on set 1.
They are:
\begin{equation}
\sigma(\pi^+ p \rightarrow \Sigma^+ K^+)
 = \frac{0.03591 (\sqrt{s}-1.688)^{0.9541}}{(\sqrt{s}-1.890)^2+0.01548}
 + \frac{0.1594 (\sqrt{s}-1.688)^{0.01056}}{(\sqrt{s}-3.000)^2+0.9412}
\hspace{0.5cm} {\rm mb},
\end{equation}
\begin{equation}
\sigma(\pi^- p \rightarrow \Sigma^- K^+)
 = \frac{0.009803 (\sqrt{s}-1.688)^{0.6021}}{(\sqrt{s}-1.742)^2+0.006583}
 + \frac{0.006521 (\sqrt{s}-1.688)^{1.4728}}{(\sqrt{s}-1.940)^2+0.006248}
\hspace{0.5cm} {\rm mb},
\end{equation}
$$
\sigma(\pi^+ n \rightarrow \Sigma^0 K^+)\hspace{2em}  {\rm and} \hspace{26em}
$$
\begin{equation}
\sigma(\pi^0 n \rightarrow \Sigma^- K^+)
= \frac{0.05014 (\sqrt{s}-1.688)^{1.2878}}{(\sqrt{s}-1.730)^2+0.006455}
\hspace{0.5cm} {\rm mb}, \hspace{12em}
\end{equation}
\begin{equation}
\sigma(\pi^0 p \rightarrow \Sigma^0 K^+)
 = \frac{0.003978 (\sqrt{s}-1.688)^{0.5848}}{(\sqrt{s}-1.740)^2+0.006670}
 + \frac{0.04709 (\sqrt{s}-1.688)^{2.1650}}{(\sqrt{s}-1.905)^2+0.006358}
\hspace{0.5cm} {\rm mb},
\end{equation}
\\
where, all parametrizations given above should be understood
to be zero below threshold for $\sqrt{s} \leq 1.688$ GeV.
These parametrizations are especially useful for kaon production simulation
codes for those channels where no experimental data are available.

To summarize, we studied the $\pi N \rightarrow \Sigma K$
reactions focusing on the role of the $\Delta(1920)$ resonance,
and presented for the first time theoretical calculations.
The results obtained
show that the isospin $I = 3/2$ $\Delta(1920)$ resonance
has a significant role to distinguish
the enegy dependence of the total cross sections for
differerent channels of the reactions $\pi N \rightarrow \Sigma K$.
We gave parametrizations of the total cross sections for each
separate channel obtained for set 1 parameters, where
the coupling constants for the $\Delta(1920)$ resonance and
$g_{K^*(892) K \pi}$ are
fitted to reproduce $\pi^+ p \rightarrow \Sigma^+ K^+$ and
other coupling constants are determined from
decay ratios. The good agreement obtained with set 1 parameters
and the failure of the set 2 parameters might indicate that
there exist other resonances
with masses around 1.9 GeV and isospin $I = 3/2$, which
contribute to the $\pi^+ p \rightarrow \Sigma^+ K^+$ channel to
explain the total cross section, and reproduce simaltaneously
also the second peak of $\pi^- p \rightarrow \Sigma^- K^+$.
\vspace{2cm}

\noindent {\bf Acknowledgement:} The authors express their thanks to
Prof. R. Vinh Mau and Prof. H. M\"uther for useful discussions.
We thank also Prof. K. W. Schmid who provided us the
code used to adjust the parameters in the present
study.

\newpage
\begin{table}
\caption{Calculated and fitted coupling constants}
\begin{center}
\begin{tabular}{cccccc}
\hline
$B^*$(resonance) &$\Gamma(MeV)$
&$\Gamma_{N \pi}(\%)$       &$g_{B^* N \pi}^2$
&$\Gamma_{\Sigma K}(\%)$    &$g_{B^* \Sigma K}^2$  \\
\hline \\
$N(1710)$     &100 &15.0 &2.57&6.0 &$4.50\times10^{+1}$\\
$N(1720)$     &150 &15.0 &$5.27\times10^{-2}$&3.5 &3.15\\
$\Delta(1920)$ (set 1)
&--&--&(1.44)&--&(3.83)\\
$\Delta(1920)$ (set 2)
&200 &12.5 &$4.17\times10^{-1}$&2.0 &1.11\\
\\
\hline
 &$f_{K^*(892) K \pi}^2$&
 &$g_{K^*(892) \Sigma N}^2$& &\\
\hline
\\
 &$6.89\times10^{-1}$& &$2.03\times10^{-1}$& &\\
 &($\Gamma$=50 MeV,&$\Gamma_{K\pi}=100$\%)& & &\\
\\
\hline
\end{tabular}
\end{center}
\end{table}
\vspace{1.5cm}
\noindent
{\bf Table 1}
\newline
The calculated or fitted coupling constants and the data used for the
calculations.
The values in brakets stand for the coupling constants
obtained by fitting to the total cross section for the $\pi^+ p
\rightarrow \Sigma^+ K^+$ reaction (set 1).
The value of $g_{K^*(892) \Sigma N}^2$ is the
fitted value to the $\pi^+ p \rightarrow \Sigma^+ K^+$ channel
when zero tensor coupling for $K^*(892) \Sigma N$ interaction ($\xi = 0$) is
applied and the  rescaled coupling constants for $\Delta(1920)$ are used.

\newpage
\noindent
{\bf Figure captions}
\vspace{1.5cm}

\noindent
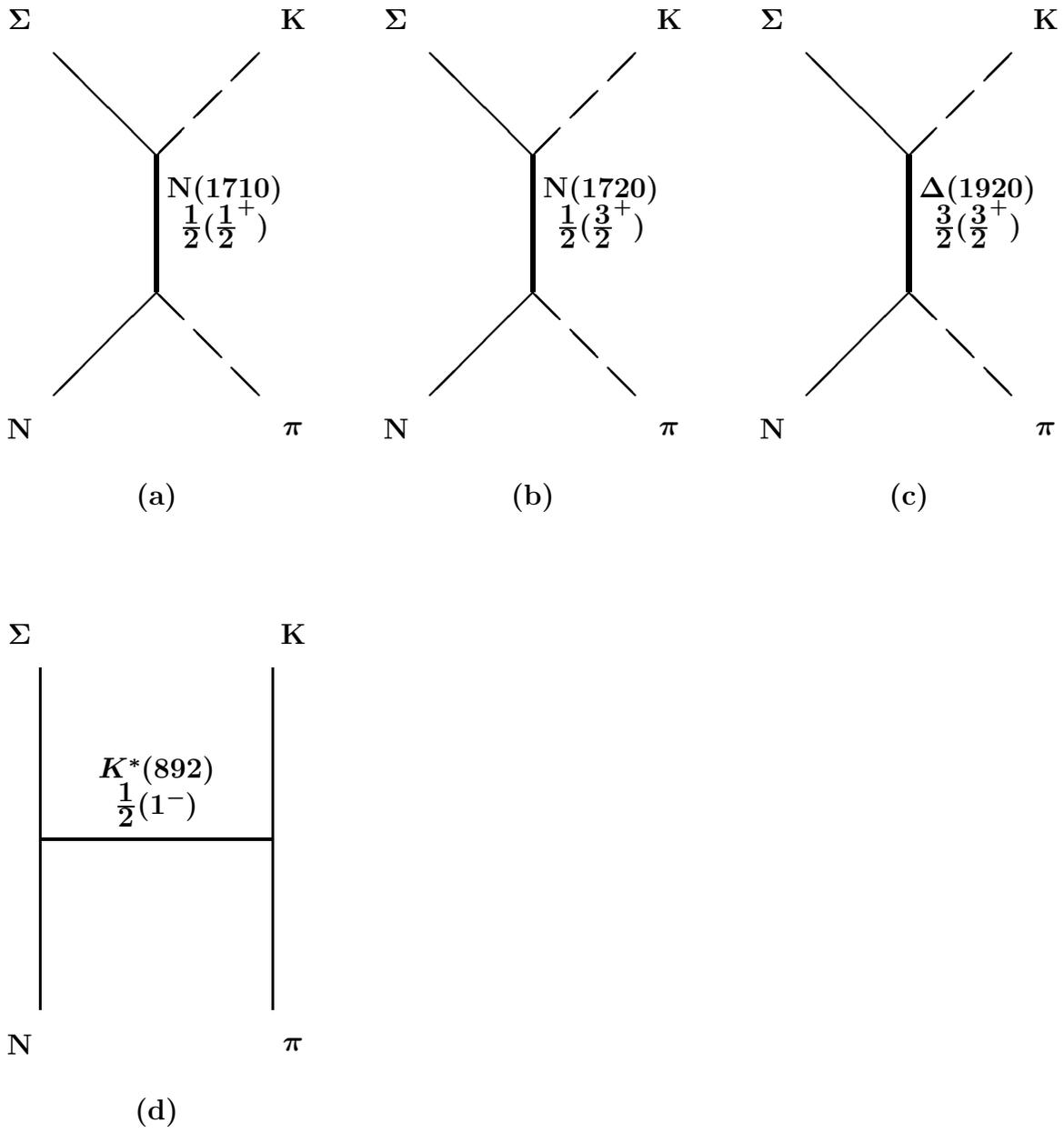
{\bf Fig. 1}
\newline
The processes contributing to the $\pi N \rightarrow \Sigma K$
reactions. The diagrams are corresponding to the different
intermediate resonance states:
$(a):\,\, N(1710)\,I(J^P) = \frac{1}{2}(\frac{1}{2}^+)$,
$(b):\,\, N(1720)\, \frac{1}{2}(\frac{3}{2}^+)$,
$(c):\,\, \Delta(1920)\, \frac{3}{2}(\frac{3}{2}^+)$ and
$(d):\,\,$t-channel $K^*(892)$ exchange, respectively.
\vspace{1cm}

\noindent
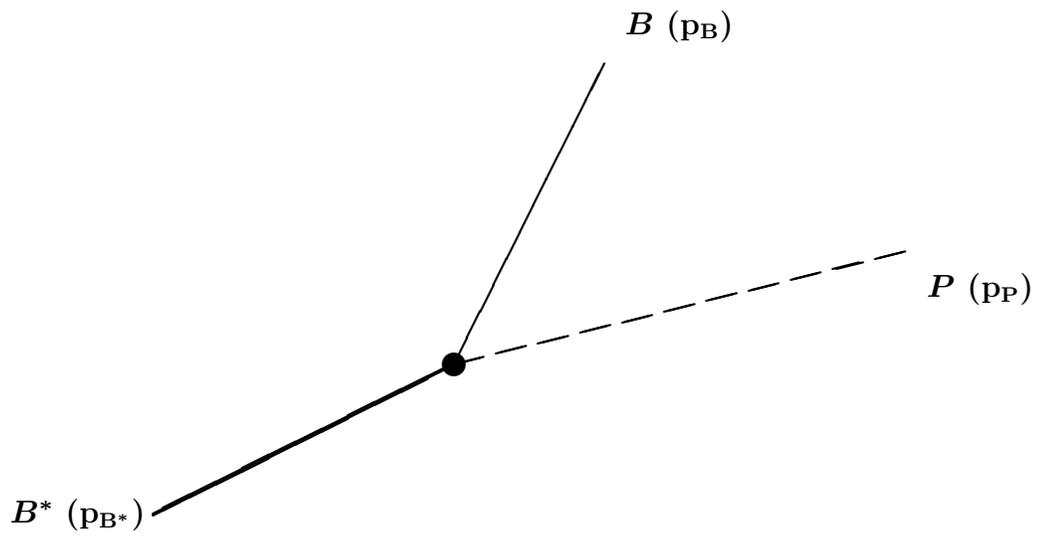
{\bf Fig. 2}
\newline
The formation or decay vertex of the resonance $B^*$ appearing
in the amplitudes eqs. (\ref{ma}) - (\ref{mc}).  Here
$B^*$ stands for the resonances $N(1710)$, $N(1720)$
and $\Delta(1920)$. $B$ stands for the baryons $N$ and
 $\Sigma$. $P$ stands for the pseudoscalar mesons
$\pi$ and $K$.
The variable given in each braket stands for the four momentum of
each particle. In the center-of-mass frame of the baryon $B$
and the pseudoscalar meson $P$, they are
$p_{B^*}^\mu = (\sqrt{s}, \vec{0}),\,\, p_B^\mu =
(E_B, \vec{p}_B)$ and $p_P^\mu = (E_P, - \vec{p}_B)$ with
$|\vec{p_B}| = q(\sqrt{s},m_B,m_P)$ defined by eq. (\ref{qdeff}).
\vspace{1cm}

\noindent
{\bf Fig. 3 (a)}
\newline
The calculated total cross sections for the
$\pi^+ p \rightarrow \Sigma^+ K^+$
($\pi^- n \rightarrow \Sigma^- K^0$) reaction.
The solid line and the dashed line stand for the set 1
and set 2 parameters, respectively. Set 1 are the relevant coupling
constants for $\Delta(1920)$ and $g_{K^*(892) K \pi}$ adjusted to the
cross section $\pi^+ N \rightarrow \Sigma^+ K^+$
with the $\Delta(1920)$ as an s-channel intermediate state resonance and other
coupling constants are determined from the branching ratios.
Set 2 uses the same value of $g_{K^*(892) K \pi}$ as that of set
1 and other coupling constants determined only from the branching ratios.
\vspace{1cm}

\noindent
{\bf Fig. 3 (b)}
\newline
The calculated total cross sections for the
$\pi^- p \rightarrow \Sigma^- K^+$
($\pi^+ n \rightarrow \Sigma^+ K^0$) reaction.
See the caption of fig. 3 (a) for further explanations.
\vspace{1cm}

\noindent
{\bf Figs. 3 (c)}
\newline
The calculated total cross sections for the
$\pi^+ n \rightarrow \Sigma^0 K^+$ and
$\pi^0 n \rightarrow \Sigma^- K^+$
($\pi^0 p \rightarrow \Sigma^+ K^0$ and
$\pi^- p \rightarrow \Sigma^0 K^0$) reactions.
See the caption of fig. 3 (a) for further explanations.
\vspace{1cm}

\noindent
{\bf Figs. 3 (d)}
\newline
The calculated total cross sections for the
$\pi^0 p \rightarrow \Sigma^0 K^+$
($\pi^0 n \rightarrow \Sigma^0 K^0$) reaction.
See the caption of fig. 3 (a) for further explanations.
No experimental data are known.
\newpage
{\Large \bf Fig. 1}\\
\bf
\boldmath
\setlength{\unitlength}{1cm}
\begin{picture}(11,20) \thicklines
\put(0,11){\makebox(1,1){N}}
\put(0,17){\makebox(1,1){$\Sigma$}}
\put(2,10){\makebox(1,1){(a)}}
\put(4,11){\makebox(1,1){$\pi$}}
\put(4,17){\makebox(1,1){K}}
\put(3,14.5){\makebox(1,1){N(1710)}}
\put(3,14){\makebox(1,1)
{$\frac{\displaystyle 1}{\displaystyle 2}
(\frac{\displaystyle 1}{\displaystyle 2}^+)$}}
\put(2.5,13.5){\line(-1,-1){1.5}}
\multiput(2.5,13.5)(0.55,-0.55){3}{\line(1,-1){0.4}}
\put(2.48,13.5){\line(0,1){2}}
\put(2.5,13.5){\line(0,1){2}}
\put(2.52,13.5){\line(0,1){2}}
\put(2.5,15.5){\line(-1,1){1.5}}
\multiput(2.5,15.5)(0.55,0.55){3}{\line(1,1){0.4}}
\put(5.5,11){\makebox(1,1){N}}
\put(5.5,17){\makebox(1,1){$\Sigma$}}
\put(7.5,10){\makebox(1,1){(b)}}
\put(9.5,11){\makebox(1,1){$\pi$}}
\put(9.5,17){\makebox(1,1){K}}
\put(8.5,14.5){\makebox(1,1){N(1720)}}
\put(8.5,14){\makebox(1,1)
{$\frac{\displaystyle 1}{\displaystyle 2}
(\frac{\displaystyle 3}{\displaystyle 2}^+)$}}
\put(8,13.5){\line(-1,-1){1.5}}
\multiput(8,13.5)(0.55,-0.55){3}{\line(1,-1){0.4}}
\put(7.98,13.5){\line(0,1){2}}
\put(8,13.5){\line(0,1){2}}
\put(8.02,13.5){\line(0,1){2}}
\put(8,15.5){\line(-1,1){1.5}}
\multiput(8,15.5)(0.55,0.55){3}{\line(1,1){0.4}}
\put(11,11){\makebox(1,1){N}}
\put(11,17){\makebox(1,1){$\Sigma$}}
\put(13,10){\makebox(1,1){(c)}}
\put(15,11){\makebox(1,1){$\pi$}}
\put(15,17){\makebox(1,1){K}}
\put(14,14.5){\makebox(1,1){$\Delta$(1920)}}
\put(14,14){\makebox(1,1)
{$\frac{\displaystyle 3}{\displaystyle 2}
(\frac{\displaystyle 3}{\displaystyle 2}^+)$}}
\put(13.5,13.5){\line(-1,-1){1.5}}
\multiput(13.5,13.5)(0.55,-0.55){3}{\line(1,-1){0.4}}
\put(13.48,13.5){\line(0,1){2}}
\put(13.5,13.5){\line(0,1){2}}
\put(13.52,13.5){\line(0,1){2}}
\put(13.5,15.5){\line(-1,1){1.5}}
\multiput(13.5,15.5)(0.55,0.55){3}{\line(1,1){0.4}}
\put(0,2){\makebox(1,1){N}}
\put(0,8){\makebox(1,1){$\Sigma$}}
\put(2,1){\makebox(1,1){(d)}}
\put(4,2){\makebox(1,1){$\pi$}}
\put(4,8){\makebox(1,1){K}}
\put(2,6){\makebox(1,1){$K^*$(892)}}
\put(2,5.5){\makebox(1,1)
{$\frac{\displaystyle 1}{\displaystyle 2}({\displaystyle 1}^-)$}}
\put(0.8,3){\line(0,1){5}}
\put(4.2,3){\line(0,1){5}}
\put(0.8,5.5){\line(1,0){3.4}}
\end{picture}
\newpage
{\Large \bf Fig. 2}\\ \\ \\
\begin{picture}(13,10) (-2,4) \thicklines
\put(0,1.5){\makebox(2,1){$B^*\,\,({\rm p_{B^*}})$}}
\put(12,4.5){\makebox(2,1){$P\,\,({\rm p_P})$}}
\put(8,8){\makebox(2,1){$B\,\,({\rm p_B})$}}
\put(6,3.98){\line(-2,-1){4}}
\put(6,4){\line(-2,-1){4}}
\put(6,4.02){\line(-2,-1){4}}
\multiput(6,4)(0.56,0.14){11}{\line(4,1){0.4}}
\put(6,4){\line(1,2){2}}
\put(6,4){\circle*{0.3}}
\end{picture}
\end{document}